\documentclass [11pt]{article}

\usepackage[latin1]{inputenc}
\usepackage{authblk}
\usepackage{graphicx}
\usepackage{dcolumn}
\usepackage{bm}
\usepackage{color} 
\usepackage{epstopdf}
\usepackage{amsmath}
\usepackage[e]{esvect}
\usepackage{fullpage}
\usepackage{cite}

\newcommand{\ket}[1]{\left|#1\right\rangle}

\title{Quantum arithmetic with the Quantum Fourier Transform}
\author[1]{Lidia Ruiz-Perez \thanks{lruiper@ribera.tel.uva.es}}
\author[1]{Juan Carlos Garcia-Escartin}
\affil[1]{Dpto. de Teor\'ia de la Se\~{n}al y Comunicaciones. ETSI de Telecomunicaci\'on. Universidad de Valladolid. Campus Miguel Delibes. Paseo Bel\'en 15. 47011 Valladolid. Spain.}
\date{\today}

\begin{document}
\maketitle
\begin{abstract}
The Quantum Fourier Transform offers an interesting way to perform arithmetic operations on a quantum computer. We review existing Quantum Fourier Transform adders and multipliers and comment some simple variations that extend their capabilities. These modified circuits can perform modular and non-modular arithmetic operations and work with signed integers. Among the operations, we discuss a quantum method to compute the weighted average of a series of inputs in the transform domain. One of the circuits, the controlled weighted sum, can be interpreted as a circuit to compute the inner product of two data vectors.\\

\end{abstract}
\section{Introduction. Quantum arithmetic}
\label{intro}
The discovery of Shor's algorithm for efficient quantum factoring \cite{Sho97} awakened an interest on the quantum implementation of the modular arithmetic operations that are the building blocks of the quantum factorization circuit. Since then, there have been many proposals on how to build the required quantum modular adders, multipliers and exponentiators using a set of elementary quantum gates.

The first suggested circuits were reversible versions of known classical implementations \cite{VBE96,BCA96}. Many subsequent proposals have been improvements and modifications of reversible generalizations of the adders and multipliers of classical digital logic \cite{Gos98,CDK04,vMI05,DKR06,AAN08,TK08,TTK10,MS12,TR13,NvM14,DRG16,Bab17}.

There are also solutions with more of a ``quantum flavour'' such as tele\-por\-ta\-tion-based operations \cite{vMMN08}, measurement-based schemes on cluster states \cite{TvM10}, repeat-until-success circuits \cite{WR14} or implementations that restrict to experimentally achievable quantum operations like the nearest-neighbour interaction \cite{CvM12}. 

A particularly elegant quantum alternative is the Quantum Fourier Transform, QFT, adder of Draper \cite{Dra00} and its generalizations in a variety of QFT adders and multipliers \cite{Bea03,BBF03,PG13,MP13}. 

In this paper, we study those systems and their applications to modular and non-modular arithmetic and discuss a QFT-based circuit to compute weighted sums. The controlled version of this circuit can be used to implement inner products. 

Section \ref{phaseenc} describes the phase encoding that permits implementing arithmetic operations in the transform domain. Section \ref{adders} introduces the basic QFT adder, including modified circuits that compute non-modular additions and work with signed integers, which provides a circuit for subtraction.

Section \ref{sec_adder} analyses the implementation of QFT adders for qubits. It takes the general adder circuit and decomposes it into a group of elementary gates acting on a collection of two-level quantum systems (qubits). The implementation rests on QFT gates and controlled rotation gates. 

Section \ref{mean} proposes a modified QFT adder to compute the arithmetic mean of a list of integers. Section \ref{QFTweight} generalizes that circuit and gives a weighted adder that sums the integers of a list after multiplication with constant weights. We also study multiplication by a constant as a particular case of a weighted sum.

Section \ref{sec_mult} describes a QFT multiplier for qubits based on the combination of modified QFT adders. The QFT multiplier is adapted in Section \ref{sec_ws} to put forward a qubit implementation of a programmable weighted adder which works with arbitrary input integers and weights. The presented quantum circuit can be used to implement different inner products between two vectors. 

We conclude the paper with a discussion of the applications of these circuits in Section \ref{discussion}.

\section{The Quantum Fourier Transform and distributed phase encoding}
\label{phaseenc}
The Quantum Fourier Transform, QFT, provides an alternative way to perform arithmetic operations on a quantum computer. Consider a $d$-dimensional system with states $\ket{x}$ from the computational basis $\left\{ \ket{0}, \ket{1}, \ldots, \ket{d-1} \right\}$. In this basis, we define the $Q\!F\!T$ operation as
\begin{equation}
{Q\!F\!T}\ket{x}=\frac{1}{\sqrt{d}} \sum_{k=0}^{d-1} e^{i \frac{ 2 \pi x k }{d}}\ket{k}.
\end{equation}

The QFT allows us to encode a number $x$ in the relative phases of the states of a uniform superposition consisting in the sum of all the states $\ket{k}$ in the computational basis, each with the same $\frac{1}{\sqrt{d}}$ amplitude. 

Imagine we want to work with natural numbers from $0$ to $d-1$. One possible encoding is mapping number $x$ into state $\ket{x}$. With the QFT we can take the information into the phases $e^{i\frac{2 \pi x k }{d}}=\omega^{xk}$ that appear together with each state $\ket{k}$ of the superposition. The QFT can be interpreted as a change of basis. We call $\ket{\phi(x)}$ to the state ${Q\!F\!T}\ket{x}$ that encodes $x$ in this new transformed basis.

We can equally define an Inverse Quantum Fourier Transform operator IQFT so that
\begin{equation}
\label{IQFT}
{I\!Q\!F\!T}\ket{k}=\frac{1}{\sqrt{d}} \sum_{x=0}^{d-1} e^{-i\frac{2 \pi x k }{d}}\ket{x}.
\end{equation}

With the direct and the inverse Fourier transforms we can move back and forth between the computational basis and the phase representation. In our notation, this conversion from the phase encoding to the computational basis is written as
\begin{equation}
{I\!Q\!F\!T}\ket{\phi(x)}={Q\!F\!T}^{-1}{Q\!F\!T}\ket{x}=\ket{x}.
\end{equation}

This phase encoding is the basic common element of all existing proposals for QFT arithmetic \cite{Dra00,Bea03,BBF03,PG13,MP13}.

\section{QFT adders}
\label{adders}
QFT addition provides a simple example of how operations with the phase encoding work. Once in the transform domain, we need quantum operators that act on the distributed phases of our quantum states. The basic element in these operators is the controlled phase gate, $C\!Z$. We start from the well-known controlled Pauli $Z$ gate which, for two input qubits $\ket{x}$ and $\ket{y}$, gives
\begin{equation}
C\!Z\ket{x}\ket{y}=e^{i\pi x y }\ket{x}\ket{y}.
\end{equation}

We can generalize the gate for $d$-dimensional systems (qudits) so that
\begin{equation}
C\!Z\ket{x}\ket{y}=e^{i\frac{2\pi x y }{d}}\ket{x}\ket{y}.
\end{equation}
As it should, when $d=2$ we recover the qubit gate. 

We can also define a modified version of the controlled phase shift gate 
\begin{equation}
\label{CZF}
C\!Z^{F}\ket{x}\ket{y}=e^{i\frac{2\pi x y }{Fd}}\ket{x}\ket{y}
\end{equation}
that introduces a factor $F$ in the divisor which will be useful later. All the $C\!Z$ and $C\!Z^F$ gates we use correspond to controlled rotation gates $C\!R(\theta)$ for different rotation angles. These gates are basic building blocks in many quantum arithmetic constructions and there are multiple proposals for their implementation with different quantum information units \cite{DWS03,FEF08,NB14,HC16}.

These ingredients are enough to give a modulo $d$ adder. We can add two numbers that are originally encoded in the computational basis by taking one of them into phase encoding and then applying a controlled phase shift. The adder comes from the sequence of operations
\begin{equation}
\label{QFTadd}
{I\!Q\!F\!T}_2 \cdot  C\!Z\cdot  {\!Q\!F\!T}_2\ket{x}\ket{y}=\ket{x}\ket{x+y\pmod d}.
\end{equation}
Here and in the following equations, when we apply a quantum gate on only a subset of all the possible input states, we introduce subindices to show on which states the gates are acting. The first operation
\begin{equation}
\ket{x}\ket{y} \stackrel{\mbox{\scriptsize{QFT}}_2}{\longrightarrow} \frac{1}{\sqrt{d}} \sum_{k=0}^{d-1} e^{i\frac{ 2 \pi y k }{d}}\ket{x}\ket{k}
\end{equation}
encodes number $y$ into the phase basis. The phase gate introduces a phase shift that is equivalent to a modulo $d$ addition in that basis, so that
\begin{equation}
\frac{1}{\sqrt{d}} \sum_{k=0}^{d-1} e^{i\frac{ 2 \pi y k }{d}} \ket{x}\ket{k}\stackrel{\scriptsize{\mbox{C\!Z}}}{\longrightarrow}\frac{1}{\sqrt{d}} \sum_{k=0}^{d-1} e^{i\frac{ 2 \pi y k }{d}}  e^{i\frac{ 2 \pi x k }{d}} \ket{x}\ket{k}.  
\end{equation}

Finally, the inverse QFT takes the result back into the computational basis with 
\begin{equation}
\begin{split}
\frac{1}{\sqrt{d}} \sum_{k=0}^{d-1} e^{i\frac{2\pi (x+y) k }{d}} \ket{x}\ket{k}
\stackrel{\mbox{\scriptsize{IQFT}}_2}{\longrightarrow} \frac{1}{d}\sum_{k,l=0}^{d-1} e^{i\frac{ 2 \pi (x+y) k }{d}} e^{-i\frac{ 2 \pi  k l }{d}}  \ket{x}\ket{l} \\ = \ket{x}\ket{x+y\pmod d}.  
\end{split}
\end{equation}

The adder can be extended to any number of inputs. Imagine we have $N$ integers $x_1$, $x_2$, $\ldots$, $x_{N}$ encoded into the state $\ket{x_1}\ket{x_2}\ldots\ket{x_N}$. Then we can repeat the sum in Equation \ref{QFTadd} with the operation
\begin{equation}
\label{QFTmultiadd}
{I\!Q\!F\!T}_N\cdot   C\!Z_{1,N}  \cdots C\!Z_{N-2,N} \cdot  C\!Z_{N-1,N}\cdot   {\!Q\!F\!T}_N \ket{x_1}\ket{x_2}\ldots\ket{x_{N-1}}\ket{x_N}
\end{equation}
that produces an output state
\begin{equation}
\ket{x_1}\ket{x_2}\ldots\ket{x_{N-1}}\ket{x_1+x_2+\ldots+ x_N\pmod d}.
\end{equation}
Here, the subindices in $C\!Z_{c,t}$ give the indices of the control state, $c$, and of the target state, $t$. Each controlled phase shift adds an integer in the phase encoding. This operation uses the minimum possible number of qudits, but it can also be interesting to preserve all the input states and store their sum in an ancillary qudit. In that case, we can apply the procedure to an initial state $\ket{x_1}\ket{x_2}\ldots\ket{x_N}\ket{0}$. The result is the sum of the $N$ integers plus $0$, which gives the same result as in the compact version.

There are a few additional modifications worth noticing. First, if we want to perform arithmetic, non-modular, additions instead of modulo $d$ addition, we can always encode the data in a system of a larger dimension $d'$ where modulo $d'$ addition and regular arithmetic addition are the same for our range of values. For instance, for two integers $x$ and $y$ between $0$ and $d-1$, the sum will always stay between $0$ and $2d-2$ and a system of dimension $d'=2d-1$ will suffice. We can take an input state $\ket{x}_d\ket{y}_d\ket{0}_{2d-1}$ with systems of dimensions $d$, $d$ and $2d-1$, respectively, and use the QFT for $d'=2d-1$ and $C\!Z$ operations that can also be defined for inputs of a different size as
\begin{equation}
C\!Z \ket{x}_d\ket{y}_{2d-1}=e^{i\frac{ 2 \pi x y}{2d-1}}\ket{x}_d\ket{y}_{2d-1}.
\end{equation}
Similarly, if we sum $N$ numbers, arithmetic addition requires a system with dimension $d'=Nd-N+1$ and we need to adapt the QFT and $C\!Z$ circuits to this new dimension.

These circuits can also perform signed addition for numbers up to $d/2$. We just need to encode positive numbers $x<d/2$ into states $\ket{x}$ and negative numbers $-x$ into states $\ket{d-x}$, in both cases in the computational basis. After the QFT, positive numbers are associated to phases $e^{i\frac{ 2 \pi x}{d}}$ below $\pi$, which correspond to a phase $e^{i\frac{2\pi x k}{d}}$ accompanying each state $\ket{k}$, and negative numbers are associated to negative phases (equivalent to phases above $\pi$ for $k=1$). The QFT adder will then perform signed addition, which gives an implementation for subtraction.

\section{QFT Adder. Qubit implementation}
\label{sec_adder}
The most common implementations of quantum logic use two-dimensional systems (qubits). In this Section, we consider the addition of numbers encoded in $n$ qubits. 

We first describe a variant on Draper's QFT Adder \cite{Dra00} to compute full arithmetic additions instead of modular additions and give its implementation with elementary gates. We consider a system composed of a collection of two-level systems (qubits). 

Let $a, b$, which are integers from 0 to $2^n-1$, be the numbers to add. Let $a_1a_2\ldots a_n$ and $b_1b_2\ldots b_n$ be the binary representations of $a$ and $b$, where $a = a_12^{n-1}+a_22^{n-2} + \ldots + a_n2^0$ and $b = b_12^{n-1}+b_22^{n-2} + \ldots + b_n2^0$. Then $\ket{a} = \ket{a_1}\otimes\ket{a_2}\otimes \ldots \otimes \ket{a_n}$ and $\ket{b} = \ket{b_1}\otimes\ket{b_2}\otimes \ldots \otimes \ket{b_n}$. 

Draper's circuit first computes the quantum Fourier transform of $a$, evolving $\ket{a}$ into $\ket{\phi(a)}$:
\begin{equation}
\ket{\phi(a)}={Q\!F\!T}\ket{a}=\frac{1}{\sqrt{N}} \sum_{k=0}^{N-1} e^{i\frac{ 2 \pi a k }{N}}\ket{k},
\end{equation}
where $N = 2^n$. Then the circuit computes the sum, using the $n$ qubits that represent the number $b$ to take $\ket{\phi(a)}$ into $\ket{\phi(a+b)}$. To perform the addition, the circuit decomposes the $C\!Z$ gates presented in Section \ref{adders} into conditional rotation phase gates of the form:
\begin{equation}
R_l = \left[
\begin{array}{cc} 
1 & 0 \\
0 & e^{2\pi i \over 2^l}
\end{array}
\right].
\end{equation}
These gates are controlled by the $n$ qubits that represent the number $b$. The combined effect of all the gates is to introduce a total phase $e^{2\pi i b k\over N}$ for each state $\ket{k}$, so that the qubits containing $b$ keep the same value while the qubit register containing the QFT of $a$ now stores $\ket{\phi\left(a+b\right)}$.

\begin{figure}[ht!]
\centering
\includegraphics[width=0.8\columnwidth]{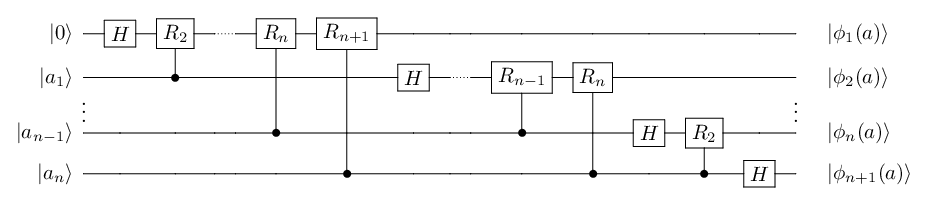}
\caption{QFT of the state $\ket{0}\ket{a}$.}\label{qft_ancillary} 
\end{figure}

We can extend the scheme to perform non-modular additions by encoding $a$ into a larger register. We represent the number $a$ using $n+1$ qubits so that $\ket{a} = \ket{0}\ket{a_1}\ket{a_2}\ldots\ket{a_n}$. The second step is computing the QFT of $\ket{a}$,
\begin{equation}
{Q\!F\!T}\ket{a}=\frac{1}{\sqrt{2^{n+1}}} \sum_{k=0}^{2^{n+1}-1} e^{i\frac{ 2 \pi a k }{2^{n+1}}}\ket{k},
\end{equation}
with the QFT circuit shown in Figure \ref{qft_ancillary}, where the states $\ket{\phi_j(a)}$ represent the $j$th qubit of the phase state $\ket{\phi(a)}$ encoding $a$. For simplicity, in this figure we have omitted the sequence of SWAP gates needed to invert the order of the output qubits \cite{NC00}. Alternatively, since we know the order of the qubits, we can connect them to the next stage in the right order.

\begin{figure}[ht!]
\centering
\includegraphics[width=0.85\columnwidth]{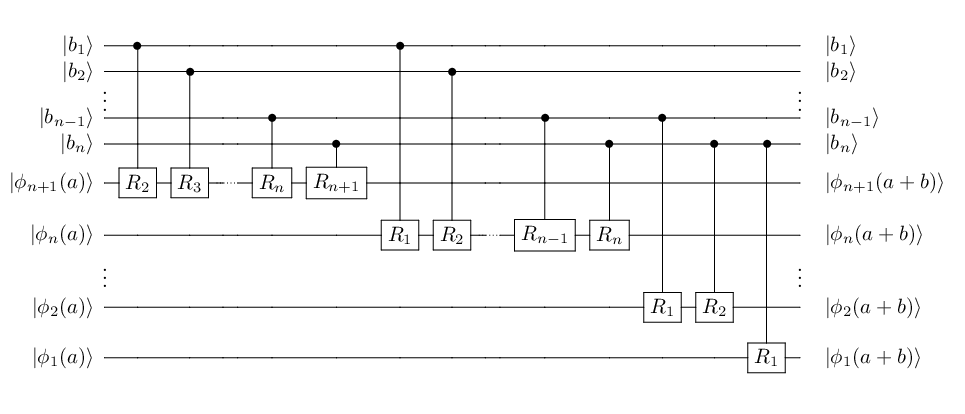}
\caption{Arithmetic (non-modular) sum in the transform domain.}\label{arithmetic_sum} 
\end{figure}

Once we have $\ket{\phi(a)}$, we add the number $b$ using controlled phase rotation gates as in Draper's scheme. We add $a$ and $b$ by applying the conditional phase rotation
\begin{equation}
e^{2\pi i \frac{(a_j+b_j)2^{n-j}k_s2^{n+1-s}}{2^{n+1}}} = e^{2 \pi i \frac{(a_j+b_j)k_s}{2^{j+s-n}}}
\end{equation}
that depends on the $j$th qubits of the representation of the numbers to be added and is applied on the $s$th qubit in the transformed register containing superpositions of states $\ket{k}=\ket{k_1}\otimes \cdots \otimes \ket{k_{n+1}}$. The gate is controlled by the $j$th qubit of $\ket{b}$ and only produces a change if $b_j=1$. We choose the conditional phase rotation gates $R_l = R_{j+s-n}$ when $j+s-n>0$. If $j+s-n \leq 0$, we are applying the phase $e^{2\pi 2^{n-j-s}i} =1$ and the state remains unaltered. The resulting circuit is shown in Figure \ref{arithmetic_sum}.

As a result, the register containing the QFT of $a$ now stores $\ket{\phi\left(a+b\right)}$. We still perform a modular addition, but, by adding an ancillary qubit to encode $a$, we avoid overflow and leave space to recover the integer addition of $a$ and $b$. When we perform the inverse QFT and go back to the computational basis, the qubit register with the sum has the correct result. This QFT adder has the minimum possible size to contain the result and needs no additional ancillary qubits, unlike some reversible adders based on classical schemes \cite{Gos98,CDK04,DKR06,TK08}. 

\section{Computing the mean with the QFT}
\label{mean}
A simple extension to the quantum adder can compute the arithmetic mean of a set of integers. We consider again $N$ integers $x_1$, $x_2$, $\ldots$, $x_{N}$ encoded into a state $\ket{x_1}\ket{x_2}\ldots\ket{x_N}$ and an ancillary $\ket{0}$ qudit. If we replace the $C\!Z$ gates in Equation (\ref{QFTmultiadd}) by the $C\!Z^{N}$ gates defined in Equation (\ref{CZF}), we have the evolution
\begin{equation}
\label{QFTmean}
\begin{split}
{I\!Q\!F\!T}_{N+1} \left(\prod_{m=1}^{N} C\!Z_{m,N+1}^N \right) {Q\!F\!T}_{N+1}\ket{x_1}\ket{x_2}\ldots\ket{x_N}\ket{0}=\\  \ket{x_1}\ket{x_2}\ldots\ket{x_N}\left|\frac{1}{N}\sum_{m=1}^{N} x_m\pmod d\right \rangle,
\end{split}
\end{equation}
which produces the desired average.

Notice that, in this case, the arithmetic mean is always equivalent to the modular addition. The mean of numbers from $0$ to $d-1$ is always between $0$ and $d-1$. However, there appears a new problem. In general, the result is not an integer and the phase to computational basis transition of the inverse QFT shown in Equation (\ref{IQFT}) does not return an integer in the computational basis. To solve this, we can expand the state space and encode the numbers in the computational basis with a fixed point representation. In Section \ref{sec_ws} we give the details of the circuit in terms of qubits for a general weighted sum. For that scenario, with $\log_2(Nd)$ qubits we recover the correct mean value. 

Alternatively, we can use the methods of phase estimation quantum algorithms which, for $d=2^m$, give the best possible $m$-bit approximation to any arbitray phase $\phi$ between 0 and 1 in a term $e^{i2\pi \phi}$ with a probability of, at least, $4/\pi^2$, which can be improved at the cost of a larger circuit \cite{CEM98}. 

\section{Weighted sums and multiplication by a constant}
\label{QFTweight}
The method of the previous Section can be modified to compute any weighted sum
\begin{equation}
\sum_{m=1}^{N}a_m x_m.
\end{equation}
We start by encoding the numbers into a state
\begin{equation}
\ket{x_1}\ket{x_2}\ldots\ket{x_N}\ket{0}
\end{equation}
and then apply the gate sequence
\begin{equation}
{I\!Q\!F\!T}_{N+1}\left( \prod_{m=1}^{N} C\!Z_{m,N+1}^{\frac{1}{a_m}}\right){\!Q\!F\!T}_{N+1}. 
\end{equation}

The resulting state 
\begin{equation}
\ket{x_1}\ket{x_2}\ldots\ket{x_N}\ket{a_1 x_1+a_2 x_2+\ldots+ a_N x_N\pmod d}
\end{equation}
returns the modulo $d$ weighted sum. If we want to obtain the non-modular weighted sum or add signed numbers, we might need to choose a different dimension for the ancillary qudit. We can recycle the encoding and circuit changes we used to modify adders in Section \ref{adders}.

A particular case happens when all the $a_m$ are positive and $\sum_m a_m=1$, like in the example of the arithmetic mean where $a_m=\frac{1}{N}$ for all $m$. Then the result is guaranteed to be between $0$ and $d-1$ and the modulo $d$ sum and the total weighted sum are always equal. However, if the $a_m$ are not integers we need either to increase the state space and use a fixed point representation or to include a phase estimation stage to recover our result.

Multiplication by a constant can be seen as a particular case of weighted sum. We can multiply two numbers $x$ and $b$, with $b$ constant and $x$ any integer from $0$ to $d-1$, using the binary decomposition of $b$. If $b$ has $n$ bits, we can write the product $bx$ as the sum
\begin{equation}
(b_{1}2^{n-1}\cdot b_{2}2^{n-2}\cdots b_{n-1}2^{1}\cdot  b_n 2^0)x=\sum_{m=1}^{n} b_{m}2^{n-m} x,
\end{equation}
which is a weighted sum with integer coefficients $a_m=b_{m}2^{n-m} $ and where all the $x_m$ are equal. Section \ref{sec_mult} describes a variation of this method that gives a QFT multiplier.

\section{QFT Multiplier}
\label{sec_mult}
\begin{figure}[ht!]
\centering
\includegraphics[width=0.8\columnwidth]{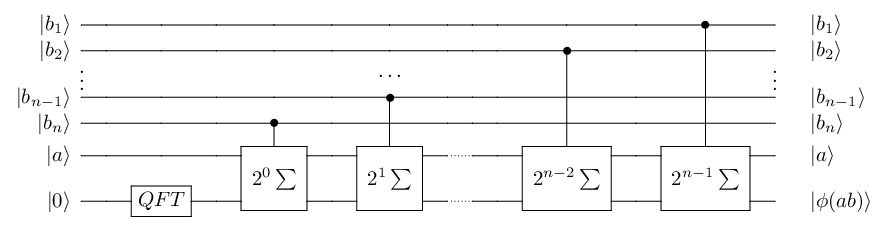}
\caption{QFT multiplier as a concatenation of controlled weighted sum blocks.}\label{qft_multiplier}
\end{figure}

We can design a quantum circuit to multiply two $n$-bit numbers by performing $n$ consecutive controlled QFT additions. The result will be a $2n$-qubit register encoding the number $a\cdot b$. The circuit is shown in Figure \ref{qft_multiplier}.

The first adder block, labelled as $2^0\Sigma$, takes as input the $n$ qubits representing a number $a$ and $2n$ qubits representing the number $0$. Before starting we prepare an initial ancillary state taking the Quantum Fourier Transform of number $0$, \textit{i.e.}, $\ket{\phi(0)}$, and then the $2^0\Sigma$ block applies a series of conditional phase rotation gates to evolve the state into $\ket{\phi(0+a)}$. The block is controlled by the least significant qubit of $\ket{b}$ so it produces the output state
\begin{equation}
\ket{a}\ket{\phi(0+b_n2^0a)}. 
\end{equation}
The next step is a second QFT adder controlled by $b_{n-1}$. Now the phase addition is scaled by a factor $2^1$ so that the output state will be
\begin{equation}
\ket{\phi(0+b_n2^0a + b_{n-1}2^1a)}. 
\end{equation}
We now proceed in a similar fashion with the remaining blocks. When the last QFT adder is applied, the output state is
\begin{equation}
\ket{\phi(0+b_n2^0a + b_{n-1}2^1a + \ldots + b_22^{n-2}a + b_12^{n-1}a)} = \ket{\phi(0+ab)} = \ket{\phi(ab)}. 
\end{equation}

The key to compute the product $a\cdot b$  is to select the proper conditional phase rotation gates to implement each QFT adder block. After computing the QFT of $0$, we obtain the output state
\begin{equation}
{Q\!F\!T}\ket{0}=\frac{1}{\sqrt{2^{2n}}} \sum_{k=0}^{2^{2n}-1} e^{i\frac{ 2 \pi 0 k }{2^{2n}}}\ket{k} = \ket{\phi(0)},
\end{equation}
where $k = k_12^{2n-1}+k_22^{2n-2} +  \ldots + k_{2n}2^0 = \sum\limits_{s=1}^{2n}k_s 2^{2n-s}$. In order to take $\ket{\phi(0)}$ to $\ket{\phi(0+b_j2^{n-j}a)}$, we need to use phase rotation gates controlled by $b_j$ and by each $a_i$, chosen so that they apply a phase rotation
\begin{equation}
e^{i\frac{ 2 \pi (a_i 2^{n-i}b_j2^{n-j}) k_s2^{2n-s}}{2^{2n}}} = e^{i\frac{ 2 \pi a_i b_j k_s }{2^{i+j+s-2n}}}.
\end{equation}
Therefore, we select conditional rotation gates of the form $R_l = R_{i+j+s-2n}$, where $i+j+s-2n>0$, to implement the QFT adder block controlled by $b_j$. 

In this circuit, we have chosen the size of the ancillary register so that we get the exact value of $a\cdot b$ instead of a modular multiplication. We can vary the size of the ancillary register and modify the $R_l$ gates accordingly to obtain any desired modular multiplication in moduli that are powers of two of the size of the ancillary register.

\section{Controlled Weighted Sum}
\label{sec_ws}
Using the same methods, we can implement a quantum circuit to compute the weighted sum
\begin{equation}
\sum\limits_{m=1}^{N}a_mx_m
\end{equation}
for any combination of input weights $a_m$ and numbers $x_m$. The weights and input integers can be in a superposition of different values. To build such a circuit we can use an architecture similar to the QFT multiplication block introduced in Section \ref{sec_mult}. 

Each qubit of $a_m$ controls how to add the contribution of each qubit of $x_m$. If we directly use the circuit of Figure \ref{qft_multiplier}, we compute the weighted sum for integer weights. However, the discrete weights $a_m$ can be adjusted to any range of interest simply by introducing the appropriate factor in the corresponding $C\!Z^{F}$ gates. We define a precision variable $p$ so that the weights $a_m$ are the integers represented by each binary string encoded in the weight qubits divided by $2^p$. Each input weight in the computational basis can be interpreted as a fixed point binary number $\ket{a}=\ket{b_1\cdots b_{q-p}.b_{q-p+1}\cdots b_{q}}$. These non-integer values are encoded into the phase representation and when, at the end of the computation, we take the inverse QFT we recover the correct result in the computational basis in the corresponding fixed point encoding. If we use $q$ qubits to store each weight $a_m$, we can obtain weights $a_m$ in a range $0\leq a_m \leq 2^{q-p}-2^{-p}$ with a precision $2^{-p}$. The values of $p$ and $q$ can be adjusted to define any desired range of values with the required precision.

\begin{figure}[ht!]
\centering
\includegraphics[width=\columnwidth]{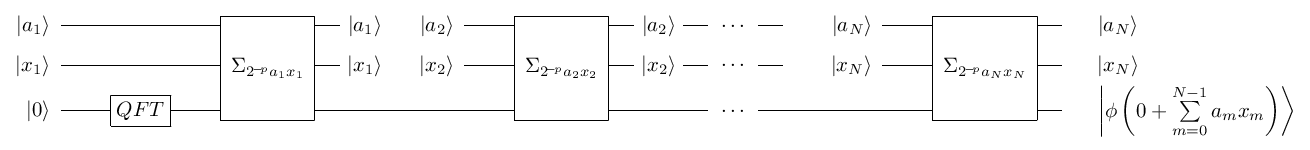}
\caption{Controlled weighted sum (block diagram).}\label{weighted_add} 
\end{figure}
The circuit can be implemented using $N$ modified versions of the QFT multiplication blocks as shown in Figure \ref{weighted_add}. We first compute the quantum Fourier Transform of $\ket{0}$. Then we apply the first multiplication block, which takes the input state 
\begin{equation}
\ket{\mathbf{a_1}}\ket{\mathbf{x_1}}\otimes \cdots \otimes \ket{a_{N}}\ket{x_{N}}\ket{\phi(0)}
\end{equation}
and returns the output state
\begin{equation}
\ket{\mathbf{a_1}}\ket{\mathbf{x_1}}\otimes \cdots \otimes \ket{a_{N}}\ket{x_{N}}\ket{\phi(0+\mathbf{a_1x_1})}.
\end{equation}
The second multiplier acts in a similar manner, taking the input state
\begin{equation}
\ket{a_1}\ket{x_1}\ket{\mathbf{a_2}}\ket{\mathbf{x_2}}\otimes \cdots \otimes \ket{a_{N}}\ket{x_{N}}\ket{\phi(0+a_1x_1)}
\end{equation}
and returning the output state
\begin{equation}
\ket{a_1}\ket{x_1}\ket{\mathbf{a_2}}\ket{\mathbf{x_2}}\otimes \cdots \otimes \ket{a_{N}}\ket{x_{N}}\ket{\phi(0+a_1x_1+\mathbf{a_2x_2})}.
\end{equation}
After applying all the multipliers, we get the final output state
\begin{equation}
\ket{a_1}\ket{x_1}\ket{a_2}\ket{x_2}\otimes \cdots \otimes \ket{a_{N}}\ket{x_{N}}\ket{\phi(0+a_1x_1+a_2x_2+\ldots+a_{N}x_{N})}.
\end{equation}

\begin{figure}[ht!]
\centering
\includegraphics[width=\columnwidth]{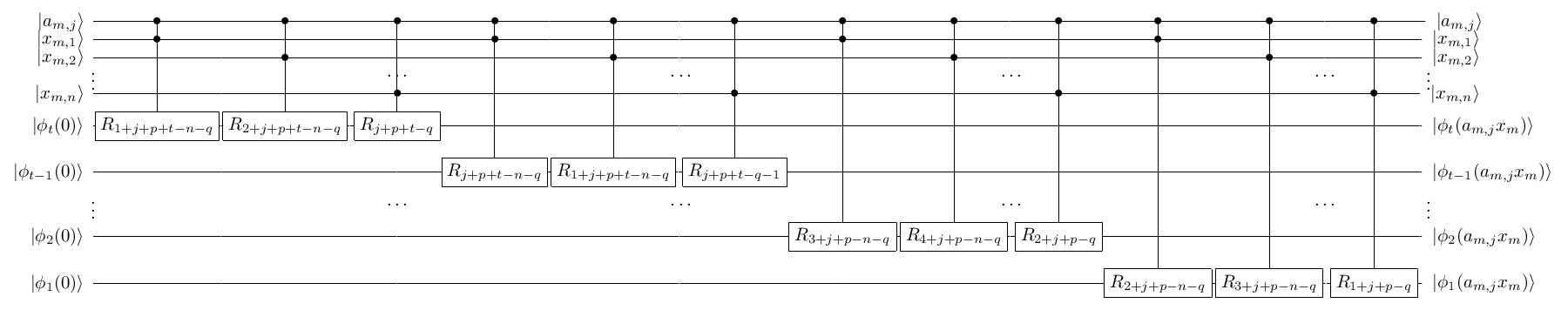}
\caption{Example block of a controlled weighted sum.}\label{weighted_sum_block} 
\end{figure}

Figure \ref{weighted_sum_block} shows an example of the gates inside each of the multiplication blocks. The Figure describes the circuit controlled by the $j$th qubit of the $m$th weight and how it controls the $R_l$ operations on the qubits of the $m$th value. The subindices are written for a system that computes the weighted sum modulo $2^t$ for weights that have $q$ bits with a precision $2^{-p}$ and input integer numbers $x_m$ with $n$ bits. If we want to find the non-modular weighted sum and there are $N$ values to be added, the ancillary register must have $t=\lceil \log_2(N 2^{q} 2^n)\rceil =\lceil (q+n)\log_2(N)\rceil$ qubits so that we can mantain the precision and there is no overflow.

The block on Figure \ref{weighted_sum_block} must be repeated for all the qubits of $\ket{a_m}$. The gate acting on the $u$th qubit of the ancillary register that is controlled by the $i$th qubit of $\ket{x_m}$ and the $j$th qubit of $\ket{a_m}$ must produce a phase shift
\begin{equation}
e^{i\frac{2\pi x_i 2^{n-i} a_j 2^{q-p-j} k_u 2^{t-u}}{2^{t}}}=e^{i\frac{2\pi x_i  a_j k_u }{2^{i+j+u+p-n-q}}},
\end{equation}
which corresponds to the conditional phase rotation gate $R_l = R_{i+j+u+p-n-q}$. 

Notice that many from these gates cancel. Any gate for which $i+j+u+p\leq n+q$ introduces a phase that is an integer multiple of $2\pi$ and can be eliminated from the final scheme.

The general weigthed sum circuit has applications in both its modular and non-modular form. The non-modular controlled weighted sum gives the dot product $\vv{a}\cdot\vv{x}$ of two vectors $\vv{a}=(a_1,\ldots,a_m)$ and $\vv{x}=(x_1,\ldots,x_m)$. The described modular operation for qubits gives an inner product in a vector space $V$ over a field consisting in a collection of $m$-tuples from $F_{p^k}^m$ where $p^k=2^n$. The $m$ elements $x_i$ and $a_i$, which are each $n$-bit binary strings, can be seen as vectors $\vv{x},\vv{a} \in V$ and the controlled weighted sum modulo $2^n$ as their inner product in $V$.

For instance, when we consider a system where the result is stored in one qubit, the resulting modular weighted sums recover the inner product modulo 2, $\sum_{i=1}^n a_i \cdot x_i \mod 2$, which appears in many quantum information protocols \cite{BV97,TS98}. The default version of our circuit, which performs modular weighted sums, generalizes this inner product to other moduli.

\section{Discussion}
\label{discussion}
The Quantum Fourier Transform offers a versatile way to perform modular and non-modular arithmetic on a quantum computer. We have discussed how QFT adders and multipliers are compact circuits for quantum
arithmetic that need no ancillary qubits and put forward a few modifications to accommodate general non-modular operations, signed numbers and different moduli. We have also discussed the qubit implementation of both QFT adders and multipliers. We can implement a QFT adder using $O(n^2)$ gates while the multiplier would need $O(n^3)$ gates for integers encoded with $n$ bits.

We have also shown that certain operations, like the arithmetic mean or any weighted average, can be implemented with the same number of gates as a basic QFT addition. If the elementary gates can be
classically programmed, this allows for a flexible quantum modular weighted sum calculator that avoids computing each weight-value product. If we sum $N$ integers, the obvious classical implementation would require $N$ multiplications and $N$ sums with a complexity $O(N(M+A))$ where $M$ and $A$ are the complexities of modular multiplication and addition respectively. For a good choice of multiplication and addition methods, the bottleneck is $M$ and we can obtain a complexity $O(N n \log^2n \log\log n)$ for Montgomery multiplication \cite{Mon85} with the Sch\"onhage-Strassen algorithm \cite{SS71}. For the our modular quantum weighted sum with integers and fixed integer weights, the complexity is comparable to $N$ sums plus the QFT and its inverse at the beginning and the end. The total complexity is $O(Nn+n^2)$. For a few long integers the dominant factor is the QFT overhead. Each sum takes $O(n)$ gates and they only become important if $N$ is of the order of the number of bits of each integer or greater. The QFT method is therefore particularly interesting when we compute the weighted sums of a large list of integers.

Additionally, we have presented a quantum circuit that computes the weight\-ed sum for both quantum weights and values. It can be implemented using $O(Ntqn)$ gates, where $n$ and $q$ are the number of bits used to encode each number $\ket{x_i}$ and each weight $\ket{a_i}$ respectively. Provided $n = q=t$, the implementation of the scheme would require a number of gates $O(N n^3)$. The circuit for $N$ numbers takes as many gates as $N$ multipliers but needs no additions. In all the cases, there is an overhead in the form of the direct and inverse QFT with $O(t^2)$ gates for modulo $2^t$ operations.

The controlled quantum weighted adder opens many applications. Optimizing weighted sums is a problem that appears in data processing and network planning among others. Many machine learning algorithms need to compute weighted sums \cite{HTJ09}. For instance, neural network training requires choosing a set of weights that minimizes a weighted sum of the samples. A quantum weighted adder that has a
uniform superposition of all the possible discrete weights for a given register size as its input can be combined with the quantum algorithm for finding the minimum \cite{DH96} to obtain a quadratic
speedup in the optimization problem. For a good enough weight precision this can be very helpful. 

While many quantum machine learning algorithms encode the weights in the probability amplitudes of a superposition \cite{RML14,SSP15}, if we want to avoid the still challenging problem of creating the initial amplitude superposition \cite{Aar15} and enter the weights and the data as states, our weighted adder gives a reversible implementation that can work with superpositions. As an alternative, we can just take direct translations of classical machine learning circuits and use methods based on Grover's algorithm \cite{Gro97} to give more modest, but reliable for any general case, quadratic speedups. 

The presented weighted sum block can also be used whenever we need an inner product inside a quantum algorithm. 

All these circuits expand the available QFT-based arithmetic operations available for quantum computers and show the potential of operations in a phase encoding.

\section*{Acknowledgements}
L. Ruiz-Perez has been funded by the FPI fellowship programme of the Spanish Ministry of Economy, Industry and Competitiveness (Grant BES-2015-074514). J.C. Garcia-Escartin has been funded by Project TEC2015-69665-R (MINECO/FEDER, UE) and Junta de Castilla y Le\'on Project No. VA089U16.

\newcommand{\noopsort}[1]{} \newcommand{\printfirst}[2]{#1}
  \newcommand{\singleletter}[1]{#1} \newcommand{\switchargs}[2]{#2#1}


\begin{thebibliography}{10}
\newcommand{\Capitalize}[1]{\uppercase{#1}}
\newcommand{\capitalize}[1]{\expandafter\Capitalize#1}

\bibitem{Sho97}
P.~W. Shor, ``Polynomial-Time Algorithms for Prime Factorization and
  Discrete Logarithms on a Quantum Computer,'' \newblock \mbox{SIAM Journal on Computing}, 26(5), ~1484 (1997).

\bibitem{VBE96}
V.~Vedral, A.~Barenco, and A.~Ekert, ``Quantum networks for
  elementary arithmetic operations,'' \newblock Physical Review A, 54(1), ~147--153 (1996).

\bibitem{BCA96}
D.~Beckman, A.~N. Chari, S.~Devabhaktuni, and J.~Preskill, ``Efficient networks for quantum factoring,'' \newblock Physical Review A, 54, ~1034--1063 (1996).

\bibitem{Gos98}
P. Gossett, ``Quantum Carry-Save Arithmetic,'' \newblock arXiv:quant-ph/9808061v2 (1998).

\bibitem{CDK04}
S.~A. Cuccaro, T.~G. Draper, S.~A. Kutin, and D.~P. Moulton, ``A
  new quantum ripple-carry addition circuit,'' \newblock arXiv:quant-ph/0410184v1 (2004).

\bibitem{vMI05}
R.~Van~Meter and K.~M. Itoh, ``Fast quantum modular exponentiation,'' \newblock Physical Review A, 71, ~052320 (2005).

\bibitem{DKR06}
T.~G. Draper, S.~A. Kutin, E.~M. Rains, and K.~M. Svore, ``A Logarithmic-depth Quantum Carry-lookahead Adder,'' \newblock Quantum Information \& Computation, 6(4), ~351--369 (2006).

\bibitem{AAN08}
J.~J. \'Alvarez-S\'anchez, J.~V. \'Alvarez-Bravo, and L.~M. Nieto, ``A quantum architecture for multiplying signed integers,'' \newblock Journal of Physics: Conference Series, 128(1), ~012013 (2008).

\bibitem{TK08}
Y. Takahashi and N. Kunihiro, ``A fast quantum circuit for addition with few qubits,'' \newblock Quantum Information \& Computation, 8(6), 636--649 (2008). 

\bibitem{TTK10}
Y. Takahashi, S. Tani and N. Kunihiro, ``Quantum addition circuits and unbounded fan-out,'' \newblock Quantum Information \& Computation, 10(9\&10), 0872--0890 (2010). 

\bibitem{MS12}
I.L. Markov and M. Saeedi, ``Constant-optimized quantum circuits for modular multiplication and exponentiation,''  \newblock Quantum Information \& Computation, 12(5\&6), 361--394 (2012). 

\bibitem{TR13}
H. Thapliyal and N. Ranganathan, ``Design of efficient reversible logic-based binary and BCD adder circuits,'' \newblock ACM Journal on Emerging Technologies in Computing Systems (JETC), 9(3), 17 (2013).

\bibitem{NvM14}
T.D. Nguyen and R. Van Meter, ``A Resource-Efficient Design for a Reversible Floating Point Adder in Quantum Computing,'' \newblock ACM Journal on Emerging Technologies in Computing Systems (JETC), 11(2), 13 (2014).

\bibitem{DRG16}
J.T. Davies, C.J. Rickerd, M.A. Grimes and D.O. Guney, ``An n-bit general implementation of Shor's quantum period-finding algorithm,'' \newblock Quantum Information \& Computation, 16(7\&8), 700--718 (2016). 

\bibitem{Bab17}
H.M.H. Babu, ``Cost-efficient design of a quantum multiplier-accumulator unit,'' \newblock Quantum Information Processing, 16(1), 30 (2017).

\bibitem{vMMN08}
R.~V. Meter, W.~J. Munro, K.~Nemoto, and K.~M. Itoh, ``Arithmetic on a Distributed-memory Quantum Multicomputer,'' \newblock Journal of Emerging Technologies in Computing Systems, 3(4), ~2:1--2:23 (2008).

\bibitem{TvM10}
A. Trisetyarso and R. Van Meter, ``Circuit Design for A Measurement-Based Quantum Carry-Lookahead Adder,'' \newblock  International Journal of Quantum Information, 08(05), 843--867 (2010).

\bibitem{WR14}
N.~Wiebe and M.~Roetteler, ``Quantum arithmetic and numerical analysis using Repeat-Until-Success circuits,'' \newblock Quantum Information \& Computation 16(1\&2), 134-178 (2016).

\bibitem{CvM12}
B.-S. Choi and R.~Van~Meter, ``A $\Theta(\sqrt{n})$-depth Quantum Adder on the 2D NTC Quantum Computer Architecture,'' \newblock Journal of Emerging Technologies in Computing Systems, 8(3), ~24:1--24:22 (2012).

\bibitem{Dra00}
T.~G. Draper, ``Addition on a quantum computer,'' \newblock  arXiv:quant-ph/0008033v1 (2000).

\bibitem{Bea03}
S.~Beauregard, ``Circuit for Shor's Algorithm Using 2n+3 Qubits,'' \newblock Quantum Information \& Computation, 3(2), ~175--185 (2003).

\bibitem{BBF03}
S. Beauregard, G. Brassard and J.M. Fernandez, ``Quantum arithmetic on Galois fields,'' \newblock arXiv:quant-ph/0301163v1 (2003).

\bibitem{PG13}
A.~Pavlidis and D.~Gizopoulos, ``Fast Quantum Modular Exponentiation Architecture for Shor's Factoring Algorithm,'' \newblock Quantum Information \& Computation, 14(7\& 8), ~649--682 (2014).

\bibitem{MP13}
C.~Maynard and E.~Pius, ``A quantum multiply-accumulator,'' \newblock Quantum Information Processing, 13(5), ~1127--1138 (2014).

\bibitem{DWS03}
J. Daboul, X. Wang and B.~C. Sanders, ``Quantum gates on hybrid qudits,'' \newblock Journal of Physics A: Mathematical and General, 36(10), 2525--2536 (2003).

\bibitem{FEF08}
I. Fushman, D. Englund, A. Faraon, N. Stoltz, P. Petroff and J. Vu\v{c}kovi\'c, ``Controlled Phase Shifts with a Single Quantum Dot,'' \newblock Science, 320(5877), 769--772 (2008).

\bibitem{NB14}
Y.~S. Nam and R. Bl\"umel, ``Robustness of the quantum Fourier transform with respect to static gate defects,'' \newblock Physical Review A, 89(4), 769--772 (2014).

\bibitem{HC16}
M. Hirose and P. Cappellaro, ``Coherent feedback control of a single qubit in diamond,'' \newblock Nature, 532(7597), 77--80 (2016).

\bibitem{NC00}
M. Nielsen and I.~L. Chuang, ``Quantum computation and quantum information,'' \newblock Cambridge University Press, Cambridge, UK (2000).

\bibitem{CEM98}
R.~Cleve, A.~Ekert, C.~Macchiavello and M.~Mosca, ``Quantum algorithms revisited,'' \newblock Proceedings of the Royal Society of London A, 454, ~339--354 (1998).

\bibitem{BV97}
E. Bernstein and U. Vazirani, ``Quantum Complexity Theory,'' \newblock SIAM Journal on Computing, 26:5, 1411-1473 (1997).

\bibitem{TS98}
B.~M. Terhal and J.~A. Smolin, ``Single quantum querying of a database,'' \newblock Physical Review A, 58(3), ~1822--1826 (1998).

\bibitem{Mon85}
P.~L. Montgomery, ``Modular Multiplication  Without  Trial  Division,'' \newblock Mathematics of Computation, 44(170),  519--521 (1985).  

\bibitem{SS71}
A. Sch{\"o}nhage and V. Strassen, ``Schnelle Multiplikation gro{\ss}er Zahlen,'' \newblock Computing, 7(3), 281--292 (1971).

\bibitem{HTJ09}
T. Hastie, R. Tibshirani and J. Friedman, ``The Elements of Statistical Learning. Data Mining, Inference, and Prediction,'' \newblock Springer Series in Statistics, Springer, New York, (2009).

\bibitem{DH96}
C.~{D\"{u}rr} and P.~{Hoyer}, ``A Quantum Algorithm for Finding the Minimum,'' \newblock eprint arXiv:quant-ph/9607014 (1996).

\bibitem{RML14}
P. Rebentrost, M. Mohseni and S. Lloyd, ``Quantum Support Vector Machine for Big Data Classification,'' \newblock Physical Review Letter, 113(13), 130503 (2014).

\bibitem{SSP15}
M. Schuld, I. Sinayskiy and F. Petruccione, ``An introduction to quantum machine learning,'' \newblock Contemporary Physics, 56(2), 172--185 (2015).

\bibitem{Aar15}
S. Aaronson, ``Read the fine print,'' \newblock Nature Physics, 11(4), 291--293 (2015). 

\bibitem{Gro97}
L.~K. Grover, ``Quantum Mechanics Helps in Searching for a Needle in a Haystack,'' \newblock Physical Review Letters, 79(2), 325--328 (1997).
\end{thebibliography}
\end{document}